\documentclass{kluwer}    

\newdisplay{guess}{Conjecture}
\usepackage[]{graphicx}
\begin{document}
\begin{article}

\begin{opening}         
\title{Massive Disks in Low Surface Brightness Galaxies} 
\author{Burkhard \surname{Fuchs}}  
\runningauthor{Burkhard Fuchs}
\runningtitle{Massive Disks in LSBGs}
\institute{Astronomisches Rechen--Institut, M\"onchhofstr.~12-14, 69120
Heidelberg, Germany}

\begin{abstract}
An update of the set of low surface brightness galaxies is presented which can 
be used to set constraints on the otherwise ambiguous decompositions of their
rotation curves into contributions due to the various components of the
galaxies. The selected galaxies show all clear spiral structure and arguments
of density wave theory of galactic spiral arms are used to estimate the
masses of the galactic disks. Again these estimates seem to indicate that the
disks of low surface brightness galaxies might be much more massive than 
currently thought. This puzzling result contradicts stellar population 
synthesis models. This would mean also that low surface brightness galaxies 
are not dominated by dark matter in their inner parts.
\end{abstract}

\end{opening}           

\section{Introduction}  
In a previous paper (Fuchs 2002) I have described how arguments of density wave
theory of galactic spiral arms can be used to set constraints on the otherwise
ambiguous decomposition of the rotation curves of low surface brightness
galaxies (LSBGs). For this purpose galaxies were selected which show clear
spiral structure. These came mainly from the set of LSBGs for which 
high--resolution rotation curves have been published by McGaugh et al.~(2001). 
The same authors (de Blok et al.~2001) have also constructed 
dynamical models of the galaxies. The observed rotation curves were modeled as 
\begin{equation}
v_c^2(R)=v_{\rm c, bulge}^2(R)+v_{\rm c, disk}^2(R)+v_{\rm c, is\,gas}^2(R)
+v_{\rm c, halo}^2(R)\,,
\end{equation}
where $v_{\rm c, bulge}$, $v_{\rm c, disk}$, $v_{\rm c, is\,gas}$, and 
$v_{\rm c, halo}$ denote
the contributions due to the bulge, the stellar disk, the interstellar gas, and
the dark halo, respectively. De Blok et al.~(2001) provide actually
for each galaxy several models, one with zero bulge and disk mass, one model
with a `reasonable' mass--to--light ratio, and a `maximum--disk' model with
bulge and disk masses at the maximum allowed by the data. All fit the data
equally well. Applying the density wave theory argument I confirmed essentially
the maximum--disk models. This result is puzzling because the mass--to--light 
ratios of these models are unaccountably high in view of stellar population 
synthesis modeling of LSBGs (cf.~Bell \& de Jong 2001). On the other hand,
this might indicate that LSBGs are less dark matter dominated than currently 
thought.

At the time of writing of their paper there was no surface photometry
available for some of the LSBGs in the set of McGaugh et al.~(2001),
so that no dynamical models could be constructed. I have inspected the images 
of these galaxies and found four galaxies, ESO\,14--40, ESO\,206--140, 
ESO\,301--120, and ESO\,425--180, which can be used for the present purpose as 
well (cf.~Fig.~1), and contrived to obtain surface photometry of the galaxies.
\begin{figure}
\centerline{\includegraphics[width=6cm]{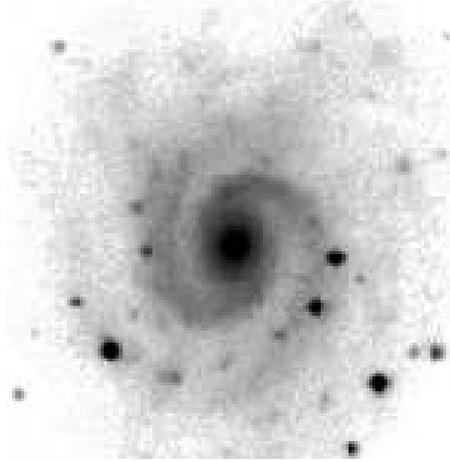}}
\caption{Image of ESO\,206--140 reproduced from Beijersbergen et al.~(1999).}
\end{figure}

\section{Update of the Sample of LSBGs}
In the meantime calibrated surface photometry has been published by
Beijersbergen et al.~(1999) for ESO\,14--40, ESO\,206--149,
and ESO\,425--180. For  ESO\,301--120 C.~M\"ollenhoff has kindly
provided uncalibrated surface photometry by reducing the blue image of the galaxy
retrieved from the Digitized Sky Survey (ESO) with his two--dimensional bulge and
disk fitting code (M\"ollenhoff \& Heidt 2001). The reliability of the
bulge and disk parameters determined this way was carefully checked with
LSBGs for which calibrated surface photometry is available. 
\begin{table}[htb]
\caption[]{Extended Sample of LSBGs.}
\begin{tabular}{lccccc}\hline
   & $v_{\rm max}$ & $v_{\rm disk,max}$ & $M_{\rm disk}$ & $ M_{\rm is gas}$ &
    $M/L_{\rm R}$\\ \hline \hline
F\,568-1    &  140 & 115 & 42 & 3 & 14 \\
F\,568-3    &  100 & 87 & 18 & 2 & 7 \\
F\,568-6    &  300 & 260 & 740 & 50 & 11 \\
F\,568-V1   &  113 & 89 & 15 & 2 & 16 \\
UGC\,128    &  131 & 102 & 43 & 9 & 4 \\
UGC\,1230   &  103 & 77 & 16 & 8 & 6 \\
UGC\,6614   &  210 & 160 & 130 & 35 & 8 \\ 
ESO\,14-40  &  272 & 205 & 250 & 83$^*$ & 4 \\
ESO\,206-140 & 119 & 93 & 27 & 10$^*$ & 4 \\
ESO\,302-120 & 89 & 70 & 11 & 11$^*$ & 1.7 \\ 
ESO\,425-180 & 140 & 120 & 62 & 38$^*$ & 2.4 \\ \hline
            & km/s & km/s & $10^9 M_\odot$ & $10^9 M_\odot$ &
	    $ M_\odot/L_{R\odot}$\\ \hline 
\end{tabular}
\end{table}
The rotation curves of the four ESO galaxies could be then interpreted 
following the prescription described in Fuchs (2002). The principal idea is that
in galactic disks spiral arms are preferentially amplified with azimuthal wave
lengths of the order of the critical wave length (cf.~Fuchs 2002 and
references therein). From the observed number of spiral arms one can thus 
estimate the critical wave lengths and from these the surface densities of 
the disks. This constrains the decomposition of the rotation curves
considerably. The resulting dynamical models of the LSBGs are summarized 
in Table 1. The second column gives the peaks of the observed rotation curves,
while the third column gives the peaks of the combined contributions due
to the stellar disks and the interstellar gas. 
Mass estimates of the combined star and gas disks of the LSBGs are given in
the fourth column. The masses of the gas disks are given separately in the fifth
column. The gas masses of the ESO galaxies have not been measured, but 
have been estimated using a relation based on the I--band luminosities of LSBGs 
derived by Schombert et al.~(2001). Similar to the other galaxies the inferred
mass--to--light ratios of the four ESO galaxies turn out to be also much 
higher than expected from population synthesis models, although they range at
the low end of the spectrum of mass--to--light ratios found for the sample of 
LSBGs studied here. A possible exception is ESO\,302--120, which appears also 
to be extremely gas rich, $M_{\rm gas} \approx M_{\rm disk}$. As can be seen
from Table 1 all dynamical models are maximum disk models. Quillen \& Pickering
(1997) have derived mass--to--light ratios for the disks of F\,568--6 and
UGC\,6614 by analyzing spiral arm induced perturbations of the velocity fields 
of the interstellar gas in the galaxies. Following this different approach they 
find also high mass--to--light ratios of the disks which agree well with the 
estimates given in Table 1.

Thus the augmented set of galaxies seems to confirm the evidence for massive 
disks in LSBGs. This does not imply, however, that the disks of LSBGs are more
massive than the disks of comparable intrinsically bright galaxies. De Blok \&
McGaugh (1996) have pointed out that the bright galaxy NGC\,2403 is very similar
to the LSBG UGC\,128. Both have about the same physical size, their rotation 
curves have the same amplitudes, and both show well developed two--armed spiral
structure. Thus the maximum disk of NGC\,2403 has the same mass as the maximum
disk of UGC\,128. Only their mass--to--light ratios are different. But the 
nature of the dim components in the disks of LSBGs remains still totally 
unclear. The relation of Schombert et al.~(2001)
implies that the gas mass of a LSBG is approximately proportional to the 
luminosity of the LSBG. Thus $M/L \propto M/M_{\rm gas}$, which might indicate 
that LSBGs with high mass--to--light ratios have consumed by star formation 
more of their interstellar gas.

\acknowledgements
I am indebted to Claus M\"ollenhoff for reducing images of LSBGs retrieved from
the Digitized Sky Survey.

\end{article}
\end{document}